\documentclass[preprint,prd,nofootinbib,tightenlines,amsmath,amssymb]{revtex4}
\usepackage{graphicx}
\usepackage{bm}
\usepackage{multirow}

\begin{document}
\baselineskip=15pt \parskip=3pt

\vspace*{3em}

\title{Probing Scotogenic Effects in Higgs Boson Decays}

\author{Shu-Yu Ho}
\affiliation{Department of Physics and Center for Theoretical  Sciences,
National Taiwan University, \\ Taipei 106, Taiwan\\}
\author{Jusak Tandean}
\affiliation{Department of Physics and Center for Theoretical  Sciences,
National Taiwan University, \\ Taipei 106, Taiwan\\}



\begin{abstract}
The recent observation of a Higgs boson at the LHC and experimental confirmation
of the nonvanishing neutrino-mixing parameter $\sin\theta_{13}$ offer important means to test
physics beyond the standard model.
We explore this within the context of the scotogenic model, in which neutrinos acquire
mass radiatively via one-loop interactions with dark matter.
Starting with a two-parameter neutrino-mixing matrix which is consistent with the latest
neutrino-oscillation data at the one-sigma level, we derive different sets of solutions
for the Yukawa couplings of the nonstandard particles in the model and use the results
to consider the Higgs decays into final states involving the new particles.
Assuming that the lightest one of them serves as fermionic cold dark matter, we show
that such decays are allowed by various experimental and theoretical constraints
to have substantial rates that are already restricted by the current LHC data.
We also look at their correlations with the Higgs decays into $\gamma\gamma$ and
$\gamma Z$.
Upcoming LHC measurements of the Higgs boson can therefore either detect
scotogenic signals or place further constraints on the model.
\end{abstract}

\maketitle

\section{Introduction}

The recent discovery of a Higgs boson with mass in the 125-126\,GeV range at
the LHC~\cite{lhc} and experimental confirmation of the neutrino-mixing parameter
$\sin\theta_{13}$ that is nonnegligible~\cite{An:2012eh} undoubtedly have far-reaching
implications for efforts to identify the nature of physics beyond the standard model~(SM).
Any realistic scenario for new physics would need to incorporate such a~particle and
account for neutrino masses and mixing angles of the right amount.
In addition, since about 22\% of the total cosmic energy density has been inferred from
astronomical observations to be attributable to matter that is nonluminous and
nonabsorbing~\cite{pdg}, the desired new-physics model should also possess at least one
candidate for dark matter~(DM).

Among the simplest possibilities accommodating the necessary ingredients is the scotogenic
model proposed by Ma~\cite{Ma:2006km}, in which neutrinos get mass radiatively via their
one-loop interactions with new particles comprising scalars and fermions, at least one of
which plays the role of~DM.
Here we explore some implications of the aforementioned experimental findings
within the context of the minimal version of this model.
In particular, identifying the newly observed Higgs boson with the Higgs boson in the model,
hereafter denoted by~$h$, and assuming the lightest one of the new fermions to be the~DM, we
consider the decays of $h$ into final states containing the nonstandard particles in the model.
Moreover, since their Yukawa couplings depend on the neutrino-mixing angles and contribute
to the decay amplitudes, it is important to adopt parametrization for the couplings that
takes into account the fact that $\sin\theta_{13}$ is not negligibly small.
Subsequently, we will demonstrate that such exotic decays of $h$ are allowed by various
experimental and theoretical constraints to have substantial rates that are already
bounded by the existing LHC data.
As upcoming measurements at the LHC will pin down the various properties of $h$ with
increasing precision, the acquired data will then either reveal hints of the new particles or
probe the model more stringently.

The organization of the paper is as follows.
The next section gives a description of the relevant Lagrangians and the neutrino mass formula.
In Section~\ref{matrices}, we write down a neutrino-mixing matrix that depends on only two
parameters and is consistent with the neutrino-oscillation data, including the measured nonzero
$\theta_{13}$, at the one-sigma level.
From the resulting neutrino-mass matrix, we derive solutions for the Yukawa couplings of
the new particles in the model.
In~Section~\ref{constraints}, we look at a number of experimental and theoretical constraints on
their couplings and masses.
Specifically, there are low-energy measurements that can limit the Yukawa couplings.
Furthermore, the Yukawa couplings belonging to the DM candidate also have to be compatible
with the observed relic abundance.
In~Section~\ref{higgsdecays}, with the parameter values satisfying the preceding constraints,
we investigate the Higgs decays into final states involving the nonstandard particles and
take into account extra constraints from the latest LHC data, including those on the Higgs decays
into \,$\gamma\gamma$\, and~\,$\gamma Z$.\,
We also examine the scotogenic impact on the correlations between these different decays.
We conclude with a summary of our results in~Section~\ref{summary}.

\section{Interactions\label{interactions}}

In its simplest version, the scotogenic model extends the minimal SM with the addition of only
a scalar doublet, $\eta$, and three neutral singlet fermions, $N_k$, all of which are
odd under an exactly conserved $Z_2$ symmetry~\cite{Ma:2006km}.
The SM particles are all even under this symmetry.
Accordingly, the lightest one of the new particles is stable and can act as~DM.
In this study, we consider the case that $N_1$ is a good candidate for
cold~DM~\cite{Kubo:2006yx}.\footnote{The possibility of $N_k$ being warm DM has also
been proposed in the literature~\cite{Sierra:2008wj}.}

The Lagrangian for the interactions of the scalar particles in this model with each other and
the standard SU(2)$_{\rm L}\times$U(1)$_Y$ gauge bosons, $\bm{W}_\rho$ and $B_\rho$, has the form
\begin{eqnarray} \label{L}
{\cal L} \,\,=\,\, ({\cal D}^\rho\Phi)^\dagger\,{\cal D}_\rho^{}\Phi \,+\,
({\cal D}^\rho\eta)^\dagger\,{\cal D}_\rho^{}\eta \;-\; {\cal V} ~,
\end{eqnarray}
where
\,${\cal D}_\rho=\partial_\rho+(i/2)g\,\bm{\tau}\!\cdot\!\bm{W}_\rho+i g_Y^{}{\cal Q}_YB_\rho$,\,
\begin{eqnarray} \label{potential}
{\cal V} &\,=&\, \mu_1^2\,\Phi^\dagger \Phi \,+\, \mu_2^2\,\eta^\dagger\eta \,+\,
\mbox{$\frac{1}{2}$}\lambda_1^{}(\Phi^\dagger \Phi)^2 \,+\,
\mbox{$\frac{1}{2}$}\lambda_2^{}(\eta^\dagger\eta)^2
\nonumber \\ && +\;
\lambda_3^{}(\Phi^\dagger \Phi)(\eta^\dagger\eta) \,+\,
\lambda_4^{}(\Phi^\dagger\eta)(\eta^\dagger \Phi) \,+\,
\mbox{$\frac{1}{2}$}\lambda_5^{}\bigl[ (\Phi^\dagger\eta)^2+(\eta^\dagger\Phi)^2\bigr] ~,
\end{eqnarray}
and, after electroweak symmetry breaking,
\begin{eqnarray}
\Phi \,\,=\, \left(\!\begin{array}{c} 0 \vspace{2pt} \\
\frac{1}{\sqrt2}(h+v) \end{array}\! \right) , \hspace{5ex}
\eta \,\,=\, \left(\!\begin{array}{c} H^+ \vspace{2pt} \\
\frac{1}{\sqrt2}({\cal S}+i{\cal P}) \end{array}\! \right) ,
\end{eqnarray}
with $g$ $(g_Y^{})$ being the usual SU(2)$_{\rm L}$ $\bigl({\rm U(1)}_Y\bigr)$ gauge coupling
constant, \,$\bm{\tau}=(\tau_1^{},\tau_2^{},\tau_3^{})$ the Pauli matrices,
${\cal Q}_Y$ the hypercharge operator, and $v$ the vacuum expectation value (VEV) of $\Phi$.
The VEV of $\eta$ is zero due to the $Z_2$ symmetry.
The masses of $\cal S$, $\cal P$, and $H^\pm$ are then, respectively,
\begin{eqnarray} & \displaystyle
m_{\cal S}^2 \,\,=\,\,
\mu_2^2 \,+\, \mbox{$\frac{1}{2}$}(\lambda_3+\lambda_4+\lambda_5)v^2 ~, \hspace{5ex}
m_{\cal P}^2 \,\,=\,\, \mu_2^2 \,+\, \mbox{$\frac{1}{2}$}(\lambda_3+\lambda_4-\lambda_5)v^2 ~,
& \nonumber \\ & \displaystyle
m_H^2 \,\,=\,\, \mu_2^2 \,+\, \mbox{$\frac{1}{2}$}\lambda_3\,v^2 ~. &
\end{eqnarray}
In our numerical analysis in Sections$\;$\ref{constraints} and~\ref{higgsdecays},
we will make the usual assumption~\cite{Kubo:2006yx} that
$\lambda_5$ is very small, \,$|\lambda_5|\ll|\lambda_3+\lambda_4|$,\, which implies that
\,$|m_{\cal S}^2-m_{\cal P}^2|=|\lambda_5|v^2\ll m_{\cal S}^2\simeq m_{\cal P}^2$.\,
From Eq.\,(\ref{L}), the couplings of $\eta$ to $h$, the photon $A$, and the $Z$ boson are
described by
\begin{eqnarray} \label{Lphieta}
{\cal L} &\,\supset&\, \bigl[ \bigl(\mu_2^2-m_{\cal S}^2\bigr){\cal S}^2 +
\bigl(\mu_2^2-m_{\cal P}^2\bigr){\cal P}^2 + 2\bigl(\mu_2^2-m_H^2\bigr)H^+H^- \bigr] \frac{h}{v}
\nonumber \\ && +\;
i e\,\bigl(H^+\,\partial^\rho H^--H^-\,\partial^\rho H^+\bigr) A_\rho^{}
\,+\, e^2\,H^+H^- A^2
\,+\, \frac{e g\,\bigl(1-2s_{\rm w}^2\bigr)}{c_{\rm w}^{}}\,H^+H^- A^\rho Z_\rho^{}
\nonumber \\ && +\;
\frac{g}{2c_{\rm w}^{}} \bigl[ {\cal P}\,\partial^\rho{\cal S}-{\cal S}\,\partial^\rho{\cal P}
\,+\, i\bigl(1-2s_{\rm w}^2\bigr) \bigl(H^+\,\partial^\rho H^--H^-\,\partial^\rho H^+\bigr)
\bigr] Z_\rho^{} ~,
\end{eqnarray}
where only terms relevant to our processes of interest are on display, \,$e=g s_{\rm w}^{}>0$\,
is the electromagnetic charge, and \,$c_{\rm w}^{}=\sqrt{1-s_{\rm w}^2}=\cos\theta_{\rm W}$\,
with the Weinberg angle $\theta_{\rm W}$.

The new singlet fermions $N_k$ are permitted to have Majorana masses
and interact with other particles according to
\begin{eqnarray} \label{LN}
{\cal L}_N^{} \,\,=\,\,
-\mbox{$\frac{1}{2}$} M_k^{}\,\overline{N_k^{\rm c}}\,P_R^{} N_k^{} \,+\,
{\cal Y}_{jk}^{} \Bigl[ \bar\ell_j^{} H^- \,-\,
\mbox{$\frac{1}{\sqrt2}$}\,\bar\nu_j^{}\,({\cal S}-i {\cal P}) \Bigr] P_R^{} N_k^{}
\;+\; {\rm H.c.} ~,
\end{eqnarray}
where \,$j,k=1,2,3$\, are summed over, the superscript c refers to charge conjugation,
\,$P_R=\frac{1}{2}(1+\gamma_5)$,\, and \,$\ell_{1,2,3}=e,\mu,\tau$.\,
Hence, writing the Yukawa couplings \,${\cal Y}_{jk}^{}=Y_{\ell_j k}$,\, we have
\begin{eqnarray} \label{yukawa}
{\cal Y} \,\,=\, \left(\begin{array}{ccc} Y_{e1} & Y_{e2} & Y_{e3} \vspace{3pt} \\
Y_{\mu1} & Y_{\mu2} & Y_{\mu3} \vspace{3pt} \\ Y_{\tau1} & Y_{\tau2} & Y_{\tau3}
\end{array}\right) .
\end{eqnarray}

In this model the light neutrinos get mass radiatively through one-loop
diagrams involving internal ${\cal S}$ or ${\cal P}$ and $N_k$.
The mass eigenvalues $m_j^{}$ of the neutrinos are given by~\cite{Ma:2006km}
\begin{eqnarray} \label{UMU} & \displaystyle
{\rm diag}\bigl(m_1^{},m_2^{},m_3^{}\bigr) \,\,=\,\,
{\cal U}^{\rm T}{\cal M}_\nu\,{\cal U} ~, & \label{YLY}
\\ & \label{Mnu} \displaystyle
{\cal M}_\nu^{} \,\,=\,\, {\cal Y}\, {\rm diag}(\Lambda_1,\Lambda_2,\Lambda_3)\,
{\cal Y}^{{\rm T}^{\vphantom{\displaystyle|}}}  ~, &
\\ & \displaystyle
\Lambda_k^{} \,\,=\,\, \frac{\lambda_5^{}\,v^2}{16\pi^2\,M_k^{}}\,
{\cal I}\Biggl(\frac{M_k^2}{m_0^2}\Biggr)^{\vphantom{\int_|^|}} \,, \hspace{5ex}
{\cal I}(x) \,\,=\,\, \frac{x}{1-x} + \frac{x^2\,\ln x}{(1-x)^2} ~, \hspace{5ex}
2m_0^2 \,\,=\,\, m_{\cal S}^2 + m_{\cal P}^2 ~, & \label{lambdak}
\end{eqnarray}
where $\,\cal U$ is the PMNS (Pontecorvo-Maki-Nakagawa-Sakata~\cite{pmns}) unitary matrix and
the expression for $\Lambda_k^{}$ is valid for \,$m_0^{}\simeq m_{\cal S}^{}\simeq m_{\cal P}^{}$.

\section{Mixing and Yukawa matrices\label{matrices}}

We express the PMNS mixing matrix as a product of two matrices involving only two mixing
angles, $\theta$ and $\varsigma$, respectively, with the latter matrix also
containing a $CP$-violating phase $\delta$.
Thus
\begin{eqnarray} \label{U}
{\cal U} &=&
\left(\begin{array}{ccc} \cos\theta & \sin\theta & 0 \vspace{1ex} \\
\frac{-1}{\sqrt2}\,\sin\theta & \frac{1}{\sqrt2}\cos\theta & \frac{1}{\sqrt2} \vspace{1ex} \\
\frac{1}{\sqrt2}\,\sin\theta & ~~ \frac{-1}{\sqrt2}\cos\theta ~~ & \frac{1}{\sqrt2}
\end{array}\right)
\left(\begin{array}{ccc} \cos\varsigma & ~~ 0 ~~ & e^{i\delta}\sin\varsigma \vspace{1ex} \\
0 & 1 & 0 \vspace{1ex} \\
-e^{-i\delta}\sin\varsigma & ~~ 0 ~~ & \cos\varsigma \end{array}\right)
\nonumber \\ &=&
\frac{1}{\sqrt2} \left(\begin{array}{ccc}
\sqrt2\;\cos\theta\,\cos\varsigma & ~ \sqrt2\;\sin\theta ~ &
\sqrt2\;e^{i\delta}\cos\theta\,\sin\varsigma \vspace{1ex} \\
-\sin\theta\,\cos\varsigma-e^{-i\delta}\sin\varsigma & \cos\theta &
-e^{i\delta}\sin\theta\,\sin\varsigma+\cos\varsigma \vspace{1ex} \\
\sin\theta\,\cos\varsigma-e^{-i\delta}\sin\varsigma  & -\cos\theta &
e^{i\delta}\sin\theta\,\sin\varsigma+\cos\varsigma \end{array}\right) .
\end{eqnarray}
The form of $\,\cal U$ with \,$\sin\theta=1/\sqrt3$\, was discussed in Ref.\,\cite{He:2011gb},
whereas the \,$\varsigma=0$\, case was treated in~Ref.\,\cite{Suematsu:2009ww}.
Both of these possibilities for $\,\cal U$ are no longer compatible with the most recent
findings, especially that $\sin\theta_{13}$ is not negligibly small~\cite{An:2012eh}.
Therefore, we will instead take
\begin{eqnarray} \label{choice}
\cos\theta\,\sin\varsigma \,\,=\,\, \sin\theta_{13}^{} ~, \hspace{5ex}
\theta \,\,\sim\,\, \theta_{12}^{} ~,
\end{eqnarray}
which lead numerically to elements of $\,\cal U$ consistent with their empirical counterparts
within one sigma.
For simplicity, we also fix \,$e^{i\delta}=1$\, in accordance with the value
\,$\delta=\bigl(300^{+66}_{-138}\bigr)^\circ$\, from the latest fit to
the global data~\cite{GonzalezGarcia:2012sz}.

Incorporating Eq.\,(\ref{U}) into the matrix diagonalization relation in Eq.\,(\ref{UMU}),
we then derive the mass eigenvalues
\begin{eqnarray}
m_1^{} &\,=&\,
\Bigl\{ Y_{e k}^2\,c_\theta^2\,c_\varsigma^2 - \sqrt2\, Y_{e k}^{} \bigl[
(Y_{\mu k}-Y_{\tau k})s_\theta^{}\,c_\varsigma^{}+(Y_{\mu k}+Y_{\tau k})s_\varsigma^{} \bigr]
c_\theta^{}\,c_\varsigma^{}
\nonumber \\ && ~\, + \mbox{$\frac{1}{2}$} \bigl[
(Y_{\mu k}-Y_{\tau k})s_\theta^{}\,c_\varsigma^{}+(Y_{\mu k}+Y_{\tau k})s_\varsigma^{} \bigr]^2
\Bigr\}_{\vphantom{\int}}\Lambda_k ~,
\nonumber \\
m_2^{} &\,=&\, \Bigl[ \mbox{$\frac{1}{2}$} (Y_{\mu k}-Y_{\tau k})^2c_\theta^2 +
\sqrt2\;Y_{e k}^{}(Y_{\mu k}-Y_{\tau k})c_\theta^{}\,s_\theta^{}
+ Y_{e k}^2\,s_\theta^2 \Bigr]_{\vphantom{\int_|^|}}^{\vphantom{\int}} \Lambda_k ~,
\nonumber \\
m_3^{}  &\,=&\,
\Bigl\{ Y_{e k}^2\,c_\theta^2\,s_\varsigma^2 + \sqrt2\, Y_{e k}^{} \bigl[
(Y_{\mu k}+Y_{\tau k})c_\varsigma^{}-(Y_{\mu k}-Y_{\tau k})s_\theta^{}\,s_\varsigma^{} \bigr]
c_\theta^{}\,s_\varsigma^{}
\nonumber \\ && ~\, + \mbox{$\frac{1}{2}$} \bigl[
(Y_{\mu k}+Y_{\tau k})c_\varsigma^{}-(Y_{\mu k}-Y_{\tau k})s_\theta^{}\,s_\varsigma^{} \bigr]^2
\Bigr\} \Lambda_k ~,   \label{numass}
\end{eqnarray}
where we have implicitly summed over \,$k=1,2,3$\, and adopted the notation
\begin{eqnarray}
c_\theta^{} \,\,=\,\, \cos\theta ~, ~~~~ s_\theta^{} \,\,=\,\, \sin\theta ~, \hspace{5ex}
c_\varsigma^{} \,\,=\,\, \cos\varsigma ~, ~~~~ s_\varsigma^{} \,\,=\,\, \sin\varsigma ~.
\end{eqnarray}
As will be seen shortly, these mass formulas can be rendered much simpler using
the relations among $Y_{\ell k}$ which have to fulfill the required vanishing of
the off-diagonal matrix elements on the right-hand side of~Eq.\,(\ref{UMU}).
Thus we arrive at the diagonalization conditions
\begin{eqnarray} \label{conditions}
0 &=& \Bigl\{
\sqrt2\; Y_{e k}\,(Y_{\mu k}-Y_{\tau k}) \bigl(c_\theta^2-s_\theta^2\bigr)c_\varsigma^{}
+ \bigl[2Y_{e k}^2-(Y_{\mu k}-Y_{\tau k})^2\bigr] c_\theta^{}s_\theta^{}\,c_\varsigma^{}
\nonumber \\ && ~ -\, \bigl(Y_{\mu k}^2-Y_{\tau k}^2\bigr) c_\theta^{}\,s_\varsigma^{}
- \sqrt2\; Y_{e k}\,(Y_{\mu k}+Y_{\tau k}) s_\theta^{}\,s_\varsigma^{}
\Bigr\}_{\vphantom{\int}}\Lambda_k ~, \nonumber \\
0 &=& \Bigl\{
\sqrt2\; Y_{e k}\,(Y_{\mu k}-Y_{\tau k}) \bigl(c_\theta^2-s_\theta^2\bigr)s_\varsigma^{}
+ \bigl[2Y_{e k}^2-(Y_{\mu k}-Y_{\tau k})^2\bigr] c_\theta^{}s_\theta^{}\,s_\varsigma^{}
\nonumber \\ && ~ +\,
\bigl(Y_{\mu k}^2-Y_{\tau k}^2\bigr) c_\theta^{}\,c_\varsigma^{}
+ \sqrt2\; Y_{e k}\,(Y_{\mu k}+Y_{\tau k}) s_\theta^{}\,c_\varsigma^{}
\Bigr\} _{\vphantom{\int}}\Lambda_k ~, \nonumber \\ \label{diagcond}
0 &=& \Bigl\{
\Bigl[ \bigl(2Y_{e k}^2-Y_{\mu k}^2-Y_{\tau k}^2\bigr) c_\theta^2
- \sqrt8\; Y_{e k}\,(Y_{\mu k}-Y_{\tau k}) c_\theta^{}s_\theta^{}
- 2 Y_{\mu k}^{}Y_{\tau k}^{}\,\bigl(1+s_\theta^2\bigr) \Bigr] c_\varsigma^{}s_\varsigma^{}
\nonumber \\ && ~ +\,
\Bigl[ \sqrt2\; Y_{e k}\, c_\theta^{}-(Y_{\mu k}-Y_{\tau k})s_\theta^{} \Bigr]
(Y_{\mu k}+Y_{\tau k}) \bigl(c_\varsigma^2-s_\varsigma^2\bigr) \Bigr\} \Lambda_k ~,
\end{eqnarray}
summation over \,$k=1,2,3$\, being again implied.
It turns out that these equations are exactly solvable for
$Y_{e k}$ and $Y_{\mu k}$ in terms of \,$Y_k\equiv Y_{\tau k}$.\,
As sketched in Appendix~\ref{yjk}, there are twenty-seven possible sets of solutions
to~Eq.\,(\ref{conditions}), but three of the sets can each produce only one nonzero mass
out of $m_{1,2,3}^{}$ in Eq.\,(\ref{numass}),
whereas another eighteen (six) of the sets can each lead to two (three) nonzero masses.

It is worth pointing out that the form of Eq.\,(\ref{Mnu}) also appears in some other models of
radiative neutrino mass, which may be generated by one-loop~\cite{1-loop} or
two-loop~\cite{2-loop} diagrams.
Hence these solutions for $Y_{\ell k}$ plus the resulting masses $m_{1,2,3}$ are also
applicable to such models, with $\Lambda_{1,2,3}$ encoding the model specifics.

The solutions in one of the eighteen sets that can each yield two nonzero masses are
\begin{eqnarray} \label{332} & \displaystyle
Y_{e i} \,\,=\,\,
\frac{\sqrt2\;c_\theta^{}\,s_\varsigma^{}\,Y_i}{s_\theta^{}\,s_\varsigma^{}+c_\varsigma^{}} ~,
\hspace{3ex} i \,\,=\,\, 1,2 ~, \hspace{5ex}
Y_{e3} \,\,=\,\, \frac{-\sqrt2\;s_\theta^{}\,Y_3}{c_\theta^{}} ~,
\nonumber \\ & \displaystyle
Y_{\mu i} \,\,=\,\, \frac{c_\varsigma^{}-s_\theta^{}\,s_\varsigma^{}}
{s_\theta^{}\,s_\varsigma^{}+c_\varsigma^{}}\,Y_i ~, \hspace{17ex}
Y_{\mu3} \,\,=\,\, -Y_3 ~. ~~~~ ~~~ &
\end{eqnarray}
These lead to the masses
\begin{eqnarray} \label{mass332}
m_1^{} \,\,=\,\, 0 ~, \hspace{5ex}
m_2^{} \,\,=\,\, \frac{2\Lambda_3^{}\,Y_3^2}{c_\theta^2} ~, \hspace{5ex}
m_3^{} \,\,=\,\,
\frac{2\bigl(\Lambda_1^{}\,Y_1^2+\Lambda_2^{}\, Y_2^2\bigr)}
{\bigl(s_\theta^{}\,s_\varsigma^{}+c_\varsigma^{}\mbox{$\bigr)^2$}} ~.
\end{eqnarray}
Setting \,$\varsigma=0$\, in the last two equations, one recovers the corresponding
expressions derived in~Ref.\,\cite{Suematsu:2009ww}.
Since it is now known experimentally that
\,$\sin^2\theta_{13}=0.0227^{+0.0023}_{-0.0024}$\,~\cite{GonzalezGarcia:2012sz},
which is not very small, the \,$\varsigma=0$\, limit is no longer a good approximation.
Particularly, as shown below, this nonzero $\theta_{13}$ corresponds to
\,$Y_{ei}\simeq0.24\,Y_{\mu i}$\, and \,$Y_{\mu i}\simeq0.82\,Y_{\tau i}$\, in~Eq.\,(\ref{332}),
compared to \,$Y_{ei}=0$\, and \,$Y_{\mu i}=Y_{\tau i}$\, in
the \,$\varsigma=0$\, case~\cite{Suematsu:2009ww}.

Hereafter, we will employ nonzero $\varsigma$ according to Eq.\,(\ref{choice}) and
focus on one of the solution sets as an example that gives three nonzero masses.
The solutions in this set are
\begin{eqnarray} \label{123} & \displaystyle
Y_{e1} \,\,=\,\, \frac{\sqrt2\;c_\theta^{}\,c_\varsigma^{}\,Y_1}
{s_\theta^{}\,c_\varsigma^{}-s_\varsigma^{}} ~, \hspace{5ex}
Y_{e2} \,\,=\,\, \frac{-\sqrt2\;s_\theta^{}\,Y_2}{c_\theta^{}} ~, \hspace{5ex}
Y_{e3} \,\,=\,\,
\frac{\sqrt2\;c_\theta^{}\,s_\varsigma^{}\,Y_3}{s_\theta^{}\,s_\varsigma^{}+c_\varsigma^{}} ~, &
\nonumber \\ & \displaystyle
Y_{\mu1} \,\,=\,\, \frac{s_\varsigma^{}+s_\theta^{}\,c_\varsigma^{}}
{s_\varsigma^{}-s_\theta^{}\,c_\varsigma^{}}\;Y_1 ~, \hspace{5ex}
Y_{\mu2} \,\,=\,\, -Y_2 ~, \hspace{5ex}
Y_{\mu3} \,\,=\,\, \frac{c_\varsigma^{}-s_\theta^{}\,s_\varsigma^{}}
{s_\theta^{}\,s_\varsigma^{}+c_\varsigma^{}}\;Y_3 ~, &
\end{eqnarray}
which yield
\begin{eqnarray} \label{mass123}
m_1^{} \,\,=\,\, \frac{2 \Lambda_1^{}\, Y_1^2}
{\bigl(s_\varsigma^{}-s_\theta^{}\,c_\varsigma^{}\mbox{$\bigr)^2$}} ~, \hspace{5ex}
m_2^{} \,\,=\,\, \frac{2 \Lambda_2^{}\, Y_2^2}{c_\theta^2} ~, \hspace{5ex}
m_3^{} \,\,=\,\, \frac{2 \Lambda_3^{}\, Y_3^2}
{\bigl(c_\varsigma^{}+s_\theta^{}\,s_\varsigma^{}\mbox{$\bigr)^2$}} ~.
\end{eqnarray}
These expressions for $m_{1,2,3}^{}$ would permit cancellations among the terms in
\,$\Delta m_{ji}^2=m_j^2-m_i^2$\, with larger $Y_k$ than would the masses in~Eq.\,(\ref{mass332}).

Numerically, we adopt for definiteness
\begin{eqnarray} \label{numchoice}
\cos\theta\,\sin\varsigma \,\,=\,\, \sqrt{0.0227} ~, \hspace{5ex} \theta \,\,=\,\, 32.89^\circ ~,
\end{eqnarray}
which translate into elements of $\,\cal U$ that are well within the one-sigma ranges of their
experimental values.
These choices imply the numbers collected in Table~\ref{ym} for the Yukawa couplings and neutrino
masses given in the previous two paragraphs in terms of $Y_k^{}$ and~\,$\Lambda_k^{}Y_k^2$.\,
After applying Eqs.$\;$(\ref{123}) and~(\ref{mass123}) to the constraints to be discussed in
the following section, we will employ the allowed parameter values to explore
several decay modes of the Higgs boson $h$.

\begin{table}[h]
\caption{Numerical values of \,$\hat Y_{\ell k}^{}=Y_{\ell k}^{}/Y_k^{}$\, and neutrino masses
in terms of \,$\tilde\Lambda_k^{}=\Lambda_k^{}Y_k^2$\, from the formulas in
Eqs.~(\ref{332})-(\ref{mass123}), with the input parameters in~Eq.\,(\ref{numchoice}).\label{ym}}
\small
\begin{tabular}{|c|ccccccccc|} \hline
~ Equations$\vphantom{\int_|^|}$ ~ & $\hat Y_{e1}$ & $\hat Y_{\mu1}$ & $\hat Y_{e2}$ & $\hat Y_{\mu2}$ &
$\hat Y_{e3}$ & $\hat Y_{\mu3}$ & $m_1^{}$ & $m_2^{}$ & $m_3^{}$ \\
\hline\hline
\,(\ref{332}), (\ref{mass332})\, &
\, 0.197 \, & \, 0.820 \, & \, 0.197 \, & \, 0.820 \, & \, $-0.915$ \, & \, $-1$ \, & \, 0 \, &
\, $2.84\,\tilde\Lambda_3$ \, & \,$1.71\bigl(\tilde\Lambda_1+\tilde\Lambda_2\bigr)^{\vphantom{|}}$\, \\
(\ref{123}), (\ref{mass123}) &
\, 3.293 \, & $-2.011$ \, & \, $-0.915$ \, & $-1$ & \, 0.197 \, & \, 0.820 \, &
\, $15.89\,\tilde\Lambda_1$ \, & \, $2.84\,\tilde\Lambda_2$ \, & \, $1.71\,\tilde\Lambda_3$ \\
\hline
\end{tabular}
\end{table}

\section{Constraints\label{constraints}}

The couplings and masses of the nonstandard particles in the scotogenic model are subject to
various constraints.
Theoretically, there are a number of restrictions on the parameters $\lambda_i$ in
the potential in~Eq.\,(\ref{potential}) and on the Yukawa couplings~${\cal Y}_{jk}$.
Vacuum stability is ensured by demanding that \,$\lambda_{1,2}>0$,\,
\,$\lambda_3>-\sqrt{\lambda_1\lambda_2}$,\,
and~\,$\lambda_3+\lambda_4\pm|\lambda_5|>-\sqrt{\lambda_1\lambda_2}$\,~\cite{Deshpande:1977rw}.
The condition of perturbativity translates into~\,$|\lambda_i|<8\pi$\,
and~\,$|{\cal Y}_{jk}|<\sqrt{4\pi}$\,~\cite{Kanemura:1999xf}.
There are additional requirements from unitarity on some combinations or functions
of~$\lambda_i$~\cite{Kanemura:1993hm,Arhrib:2012ia}.

Experimentally, there are constraints on the masses of the new scalars.
The data on $W$ and $Z$ widths and the null results of direct searches for new particles
at $e^+e^-$ colliders imply that~\cite{Arhrib:2012ia,Cao:2007rm,Pierce:2007ut}
\begin{eqnarray}
m_H^{}+m_{{\cal S},{\cal P}}^{} \,\,>\,\, m_W^{} ~, \hspace{5ex}
m_H^{} \,\,\gtrsim\,\, 70 {\rm\;GeV} ~, \hspace{5ex}
m_{\cal S}^{}+m_{\cal P}^{} \,\,>\,\, m_Z^{} ~,
\end{eqnarray}
where the last inequality is valid for
\,$|m_{\cal S}^{}-m_{\cal P}^{}|<8$\,GeV\,~\cite{Pierce:2007ut}, which
pertains to our assumption that~\,$m_{\cal S}^{}\simeq m_{\cal P}^{}$.\,
Accordingly, in our numerical analysis later on we will consider the mass regions
\,$50{\rm\,GeV}\le m_{{\cal S},{\cal P}}^{}\le120$\,GeV\,
and~\,$70{\rm\,GeV}\le m_H^{}\le 120$\,GeV.\,
These choices respect the limits on the oblique parameters $S$ and $T$
at~90\%$\;$CL (Confidence Level)~\cite{pdg,Barbieri:2006dq}.

There are also experimental constraints on the Yukawa couplings ${\cal Y}_{jk}^{}$
from several low-energy measurements and  the observed DM relic abundance,
which we address in the rest of this section.
The latest LHC data on the Higgs boson imply extra restrictions on the model,
which we will take into account in Section~\ref{higgsdecays}.

\subsection{Low-energy observables\label{leo}}

Neutrino oscillation measurements determine the differences
\,$\Delta_{ji}^2=m_j^2-m_i^2$.\,
From the latest fit to the data~\cite{GonzalezGarcia:2012sz}
\begin{eqnarray} \label{dm2}
\Delta_{21,\rm exp}^2 \,\,=\,\,
\bigl(7.50_{-0.19}^{+0.18}\bigr)\times10^{-5} {\rm\;eV}^2 ~, \hspace{5ex}
\Delta_{31,\rm exp}^2 \,\,=\,\,
\bigl(2.473_{-0.067}^{+0.070}\bigr)\times10^{-3} {\rm\;eV}^2 ~.
\end{eqnarray}
We have here assumed the normal ordering of the neutrino masses, which is preferred
to the inverted ordering by the solutions in Eq.\,(\ref{123}).
In restricting the new couplings, we will then impose
\begin{eqnarray} \label{Delta2}
31.0 \,\,<\,\, \frac{\Delta_{31}^2}{\Delta_{21}^2} \,\,<\,\, 35.0
\end{eqnarray}
based on the 90\%\,CL ranges of the numbers in~Eq.\,(\ref{dm2}).

For the masses, there are also constraints on the effective mass parameters
$\bigl\langle m_\beta^{}\bigr\rangle$ and $\bigl\langle m_{\beta\beta}^{}\bigr\rangle$
from beta decay and neutrinoless double-beta decay experiments, respectively,
and on the sum of masses, $\mbox{\large$\Sigma$}_{k\,}^{} m_k^{}$, from astrophysical and
cosmological observations.
The limits are~\cite{numasslimits}
\begin{eqnarray}
\bigl\langle m_\beta^{}\bigr\rangle \,\,=\,\,
\sqrt{\raisebox{0.5ex}{\footnotesize$\displaystyle\sum_k$}\,
\bigl|{\cal U}_{1k}^{}\bigr|^{2\,} m_k^2} \,\,<\,\, 2.1{\rm\;eV} ~, \hspace{5ex}
\bigl\langle m_{\beta\beta}^{}\bigr\rangle \,\,=\,\,
\Bigl| \raisebox{0.5ex}{\footnotesize$\displaystyle\sum_k$}\,
{\cal U}_{1k\,}^2 m_k^{}\Bigr| \,\,<\,\, 0.25{\rm\;eV} ~, \hspace{5ex}
\end{eqnarray}
and \,$\mbox{\large$\Sigma$}_{k\,}^{} m_k^{}<(0.5\mbox{-}1.5)\;$eV.\,
As it will turn out, these are less restrictive on the Yukawa couplings than
the other constraints described in this subsection.

The interactions of $H^\pm$ and $N_k$ with charged leptons give rise to
the flavor-changing radiative decay \,$\ell_j\to\ell_{i\,}\gamma$\, at one-loop order.
Such decays have been searched for, with negative results so far, including the fresh one
for \,$\mu\to e\gamma$\, reported by the MEG Collaboration~\cite{Adam:2013mnn}.
The experimental bounds on their branching ratios are~\cite{pdg,Adam:2013mnn}
\begin{eqnarray} & \label{l2l'g} \displaystyle
{\cal B}(\mu\to e\gamma)_{\rm exp} \,\,<\,\, 5.7\times10^{-13} ~, \hspace{5ex}
{\cal B}(\tau\to e\gamma)_{\rm exp} \,\,<\,\, 3.3\times10^{-8} ~,
& \nonumber \\ & \displaystyle
{\cal B}(\tau\to\mu\gamma)_{\rm exp} \,\,<\,\, 4.4\times10^{-8}  &
\end{eqnarray}
at 90\%\,CL.
Hence they put a cap on the prediction~\cite{Kubo:2006yx,Ma:2001mr}
\begin{eqnarray}
{\cal B}(\ell_j\to\ell_{i\,}\gamma) \,\,=\,\,
\frac{3\alpha\,{\cal B}(\ell_j\to\ell_{i\,}\nu\bar\nu)}{64\pi\,G_{\rm F}^2\,m_H^4}
\Bigl|\raisebox{0.5ex}{\footnotesize$\displaystyle\sum_k$}\,{\cal Y}_{ik}^{}{\cal Y}_{jk}^*\,
{\cal F}\bigl(M_k^2/m_H^2\bigr) \Bigr|^2 ~,
\end{eqnarray}
where \,$G_{\rm F}=v^{-2}/\sqrt2$\, is the Fermi constant,
\begin{eqnarray}
\alpha \,\,=\,\, \frac{e^2}{4\pi} ~, \hspace{5ex}
{\cal F}(x) \,\,=\,\, \frac{1-6x+3x^2+2x^3-6x^2\,\ln x}{6(1-x)^4} ~,
\end{eqnarray}
and numerically for ${\cal B}(\ell_j\to\ell_{i\,}\nu\bar\nu)$ we will use the central
values of their data:
\,${\cal B}(\mu\to e\nu\bar\nu)_{\rm exp}\simeq1$,\,
\,${\cal B}(\tau\to e\nu\bar\nu)_{\rm exp}=0.1783\pm0.0004$,\, and
\,${\cal B}(\tau\to\mu\nu\bar\nu)_{\rm exp}=0.1741\pm0.0004$\,~\cite{pdg}.

At one-loop level, the presence of $H^\pm$ and $N_j$ also induces a modification to
the anomalous magnetic moment $a_{\ell_i}$ of lepton $\ell_i$ given by~\cite{Ma:2001mr}
\begin{eqnarray} \label{g-2}
\Delta a_{\ell_i}^{} \,\,=\,\, \frac{-m_{\ell_i}^2}{16\pi^2m_H^2}\,
\raisebox{0.5ex}{\footnotesize$\displaystyle\sum_k$}\,|{\cal Y}_{ik}|^2\,
{\cal F}\bigl(M_k^2/m_H^2\bigr) ~.
\end{eqnarray}
The existing data on $a_{e,\mu,\tau}^{}$ and the charged-lepton masses imply that only
$a_\mu^{}$ can be significantly restrictive on the potential scotogenic effects at present.
Its most up-to-date SM and experimental values differ by nearly three sigmas,
\,$a_\mu^{\rm exp}-a_\mu^{\rm SM}=(249\pm87)\times10^{-11}$~\cite{Aoyama:2012wk}.
Accordingly, in view of the negative sign in Eq.\,(\ref{g-2}), we may require
\begin{eqnarray} \label{gmu-2}
\bigl|\Delta a_\mu^{}\bigr| \,\,<\,\, 9\times10^{-10} ~.
\end{eqnarray}
This will turn out to be complementary to Eqs. (\ref{Delta2}) and (\ref{l2l'g})
in restraining the Yukawa parameters.

\subsection{Fermionic dark matter\label{darkmatter}}

Since we have picked $N_1$ to be the lightest of the nonstandard particles
and serve as cold~DM, it needs to account for the observed cosmic relic abundance,
which therefore imposes bounds on~${\cal Y}_{k1}^{}$.
The $N_1$ annihilation cross-section $\sigma_{\rm ann}^{}$ is related to its relic
density $\Omega$ by~\cite{Kubo:2006yx,Jungman:1995df}
\begin{eqnarray} \label{omega}
\Omega \hat h^2 \,\,=\,\, \frac{1.07\times10^9\;x_f^{}{\rm\;GeV}^{-1}}
{\sqrt{g_*^{}}\;m_{\rm Pl}^{}\,\bigl[a+3(b-a/4)/x_f^{}\bigr]} ~, \hspace{5ex}
x_f^{} \,\,=\,\,
\ln\frac{0.0955\,\bigl(a+6b/x_f^{}\bigr)M_1^{}\,m_{\rm Pl}^{}}{\sqrt{g_*^{}\,x_f^{}}} ~
\end{eqnarray}
where $\hat h$ denotes the Hubble parameter,
\,$m_{\rm Pl}^{}=1.22\times10^{19}$\,GeV\,~is the Planck mass,  $g_*^{}$ is the number of
relativistic degrees of freedom below the freeze-out temperature~\,$T_f^{}=M_1^{}/x_f^{}$,
and $a$ and $b$ are defined by the expansion
\,$\sigma_{\rm ann}^{}v_{\rm rel}^{}=a+b v_{\rm rel}^2+{\cal O}\bigl(v_{\rm rel}^4\bigr)$
in terms of the relative speed $v_{\rm rel}^{}$ of the $N_1\bar N_1$ pair in their
center-of-mass frame.

The leading contributions to $\sigma_{\rm ann}^{}$ are the tree-level processes
\,$N_1^{}\bar N_1^{}\to\ell_i^-\ell_j^+$\, and \,$N_1^{}\bar N_1^{}\to\nu_i^{}\bar\nu_j^{}$\,
via exchanges of $H^\pm$ and $({\cal S,P})$, respectively, each of which proceeds from
diagrams in the $t$ and $u$ channels because of the Majorana nature of the external
neutral fermions.
We collect the expressions for their squared amplitudes in Appendix~\ref{Mann} for
\,$m_0^{}\simeq m_{\cal S}^{}\simeq m_{\cal P}^{}$.\,
If all the final-lepton masses are negligible, their combined cross-section
times $v_{\rm rel}^{}$ is
\begin{eqnarray}
\sigma_{\rm ann}^{}v_{\rm rel} \,\,=\,\, \sum_{i,j=1,2,3}
\frac{|{\cal Y}_{i1}{\cal Y}_{j1}|^2 M_1^2 v_{\rm rel}^2}{48\pi}
\Biggl[ \frac{M_1^4+m_H^4}{\bigl(M_1^2+m_H^2\bigr)^4} \,+\,
\frac{M_1^4+m_0^4}{\bigl(M_1^2+m_0^2\bigr)^4} \Biggr]
\end{eqnarray}
to second order in $v_{\rm rel}^{}$, which implies \,$a=0$\, and \,$b\neq0$.\,
The \,$m_H^{}=m_0^{}$\, limit of this formula agrees with that found in Ref.~\cite{Kubo:2006yx}.
In our numerical computation with more general masses, we employ the cross section obtained
from the squared amplitudes in~Appendix~\ref{Mann} and do not neglect the charged lepton
masses, in which case $a$ is also nonvanishing.

\begin{figure}[h] \vspace{2ex}
\includegraphics[width=86.8mm]{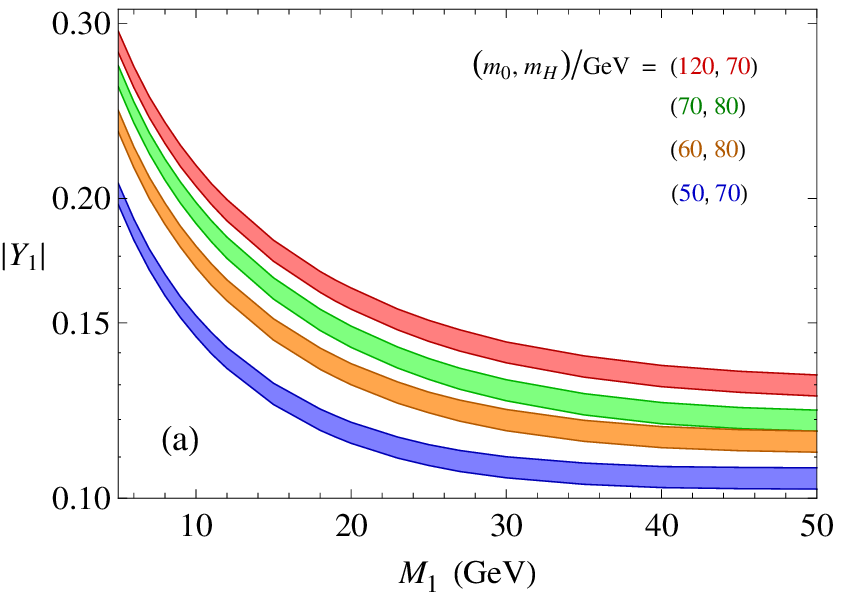} \,
\includegraphics[width=87mm]{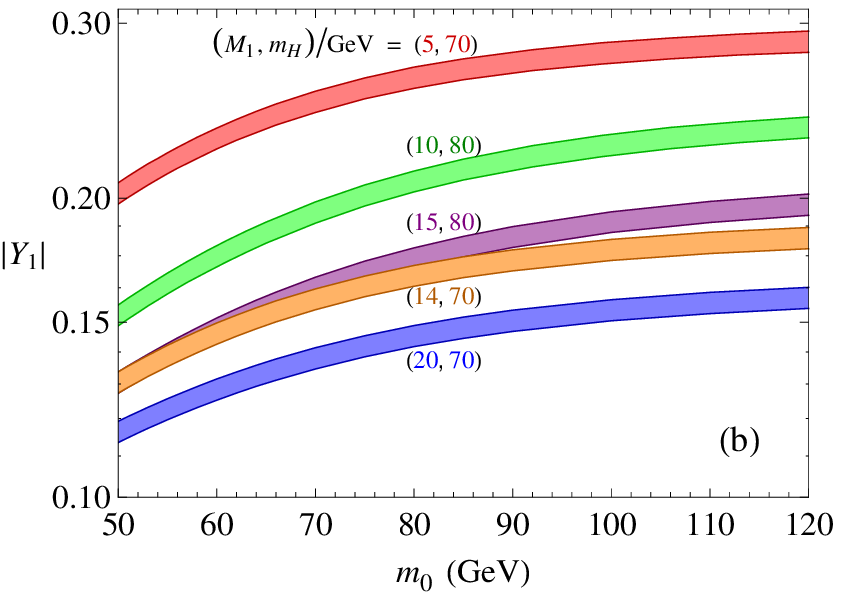}\vspace{-2ex}
\caption{Magnitude of $Y_1$ belonging to Yukawa couplings in Eq.\,(\ref{123}) versus
(a)~$M_1$ and (b)~$m_0^{}$ satisfying the relic density constraint
\,$0.101\le\Omega\hat h^2\le0.121$\, for some choices of
$\bigl(m_0^{},m_H^{}\bigr)$ and $\bigl(M_1^{},m_H^{}\bigr)$, respectively.\label{Y1plots}}
\end{figure}

Since the solutions for ${\cal Y}_{k1}$ derived in Section~\ref{matrices} are all
proportional to $Y_1$, it is the only coupling relevant to the relic density of $N_1$.
To extract $|Y_1|$ from the empirical value of $\Omega$, one can utilize Eq.\,(\ref{omega})
once the mass parameters $m_{0,H}^{}$ and $M_1^{}$ are specified.
Thus, we present in Figure~\ref{Y1plots}(a) some samples of the values of $|Y_1|$
consistent with the 90\%\,CL range of the data \,$\Omega\hat h^2=0.111\pm0.006$\,~\cite{pdg}
over \,$5{\rm\;GeV}\le M_1\le50\;$GeV\, for the solutions in Eq.\,(\ref{123})
and various sets of $m_{0,H}^{}$.\,
To offer a~different perspective on the allowed values, in Figure~\ref{Y1plots}(b)
we display $|Y_1|$ versus $m_0^{}$ for several choices of $M_1$ and $m_H^{}$.
The plot of $|Y_1|$ versus $m_H^{}$ could be roughly inferred from Figure~\ref{Y1plots}(b)
by interchanging $m_0^{}$ and $m_H^{}$, especially for~\,$m_\ell^{}\ll M_1$.\,
Evidently, the demand that $N_1$ be the leading candidate for cold DM over the mass regions of
interest can always be met by some values of~$Y_1$, but their ranges are fairly limited.

Direct-search experiments for DM, which look for signals of it colliding with nuclei,
may lead to further restraints on $Y_1$, but the existing bounds are still too weak.
Since $N_1$ can scatter off a~nucleon mainly via its one-loop $Z$-mediated axial-vector
interactions with quarks~\cite{Suematsu:2009ww,Cao:2009yy}, the process is characterized by
a spin-dependent cross-section that is relatively suppressed and in our case does not
reach~$10^{-41}{\rm\,cm}^2$.
This is more than an order of magnitude below the strictest limit to date, measured by
the XENON100 Collaboration~\cite{Aprile:2013doa}.

 \vspace{-1ex}

\section{Implications for Higgs boson decay\label{higgsdecays}}

As experimental work on the Higgs boson proceeds at the LHC with increasing precision,
the accumulated data will reveal how much the properties of the particle may deviate
from SM expectations.
The information gained will then serve to test in particular various scenarios in which
new physics can induce nonstandard decay modes of the Higgs and/or
significant modifications to its SM decay channels~\cite{exotic}.
In the scotogenic model, such effects can arise at tree and loop levels, to which we now turn.
Using the parameter space allowed by the constraints discussed above, we first look at
Higgs decays into final states containing the new particles.
As mentioned earlier, in our numerical work below we assume
\,$50{\rm\,GeV}\le m_0^{}\le120$\,GeV\, and \,$70{\rm\,GeV}\le m_H^{}\le120$\,GeV.\,

Since collider data imply \,$2m_0^{}>m_Z^{}$,\, our mass ranges of interest include
\,$m_Z^{}<2m_0^{}\le m_h^{}$,\, in which case the decay channels
\,$h\to {\cal S}{\cal S},{\cal P}{\cal P}$\, are open and may be important~\cite{Cao:2007rm}.
From~Eq.\,(\ref{Lphieta}), we obtain the amplitudes for these modes at tree level to be
\begin{eqnarray} \label{h2ss}
{\cal M}_{h\to {\cal S}{\cal S}}^{} \,\,\simeq\,\, {\cal M}_{h\to {\cal P}{\cal P}}^{}
\,\,\simeq\,\, \frac{2\bigl(m_0^2-\mu_2^2\bigr)}{v} ~.
\end{eqnarray}
Their combined rate is
\begin{eqnarray} \label{Gh2ss}
\Gamma(h\to\eta^{\scriptscriptstyle0}\eta^{\scriptscriptstyle0}) \,\,=\,\,
\Gamma(h\to{\cal S}{\cal S})+\Gamma(h\to{\cal P}{\cal P}) \,\,\simeq\,\,
\frac{\bigl(m_0^2-\mu_2^2\bigr)^2}{4\pi\,m_h^{}v^2}\sqrt{1-\frac{4m_0^2}{m_h^2}} ~.
\end{eqnarray}
The decay products would be invisible for ${\cal S}$ and ${\cal P}$ lighter than $N_k$.
Otherwise, ${\cal S}$ and ${\cal P}$ will decay into \,$\nu_j^{}N_k^{}$\, and,
if kinematically possible, $N_k$ will subsequently decay into
\,$\ell_j^+\ell_{j'}^-N_{k'}^{}$ $\bigl(\bar\nu_j^{}\nu_{j'}^{}N_{k'}^{}\bigr)$ through
$H^\pm$ $({\cal S,P})$ exchange.

In the mass range \,$m_h^{}/2<m_0^{}<m_h^{}$,\, these two-body decays no longer happen,
and so the dominant modes with $\cal S$ and $\cal P$ in the final states are the three-body
decays \,$h\to ({\cal S},{\cal P})\nu_j^{}N_k^{}$\, if kinematically permitted.
From Eqs.$\;$(\ref{Lphieta}) and~(\ref{LN}), we obtain the tree-level amplitude
\begin{eqnarray} \label{h2snN}
{\cal M}_{h\to {\cal S}({\cal P})\nu_j^{}\bar N_k^{}} \,\,=\,\, (-i)\frac{\sqrt2}{v}\,
\frac{\bigl(\mu_2^2-m_0^2\bigr)\,{\cal Y}_{jk}^{}\,\bar\nu_j^{}P_R^{}N_k^{}}
{m_0^2-\bigl(p_\nu^{}+p_N^{}\mbox{$\bigr)^2$}}
\end{eqnarray}
and an analogous expression for \,$h\to {\cal S}({\cal P})\bar\nu_j^{}N_k^{}$.\,
They contribute to the combined rate
\begin{eqnarray} \label{Gh2enN}
\Gamma(h\to\eta^{\scriptscriptstyle0}\nu N) \,\,=\,\,
\sum_{\hat\eta=\cal S,P}\,\sum_{j,k=1,2,3} \bigl[
\Gamma\bigl(h\to\hat\eta\,\nu_j^{}\bar N_k^{}\bigr)+
\Gamma\bigl(h\to\hat\eta\,\bar\nu_j^{}N_k^{}\bigr) \bigr] ~.
\end{eqnarray}

For decays with $H^\pm$ in the final states, since collider data restrain their mass to
be~\,$m_H^{}\mbox{\footnotesize\;$\gtrsim$\;}70$\,GeV,\, we only have the three-body
modes \,$h\to H^+\ell_j^-\bar N_k^{}$\, and their charge-conjugated counterparts,
provided that \,$m_H^{}+m_\ell^{}+m_N^{}<m_h^{}$.\,
As in the neutral case, we find
\begin{eqnarray} \label{h2HlN}
{\cal M}_{h\to H^+\ell_j^-\bar N_k^{}}^{} \,\,=\,\, \frac{2}{v}\,
\frac{\bigl(m_H^2-\mu_2^2\bigr)\,{\cal Y}_{jk}^{}\,\bar\ell_j^{}P_R^{}N_k^{}}
{m_H^2-\bigl(p_\ell^{}+p_N^{}\mbox{$\bigr)^2$}}
\end{eqnarray}
and similarly for \,$h\to H^-\ell_j^+N_k^{}$.\,
They lead to the rate
\begin{eqnarray} \label{Gh2HlN}
\Gamma(h\to H\ell N) \,\,=\,\, \sum_{j,k=1,2,3} \bigl[
\Gamma\bigl(h\to H^+\ell_j^-\bar N_k^{}\bigr)+
\Gamma\bigl(h\to H^-\ell_j^+N_k^{}\bigr) \bigr] ~,
\end{eqnarray}
which is identical in form to Eq.\,(\ref{Gh2enN}).

After fixing $Y_1$ to the range fulfilling the relic density requirement as
in Section~\ref{darkmatter}, we scan all the relevant parameter space of the model
subject to the constraints described in Section~\ref{leo}.
We find that, for the allowed coupling values, these two- and three-body decay channels
of the Higgs can have enlarged rates.
We illustrate this in Table$\;$\ref{numbers} for different sets of $m_{0,H}^{}$,
$\mu_2^{}$, and $M_{1,2,3}$, employing the Higgs mass \,$m_h^{}=125.5$\,GeV,\,
compatible with the most recent measurements~\cite{higgsmasses},
and SM Higgs total width~\,$\Gamma_h^{\rm SM}=4.14\;$MeV\,~\cite{lhctwiki}.
The numbers we have selected for $m_{0,H}^{}$, $M_1$, and $Y_1$ can be seen to correspond to
some of the points inside the colored bands on one or both graphs in~Figure$\;$\ref{Y1plots}.
In the tenth column of this table, we list the branching ratio
\,${\cal B}_{\scriptscriptstyle{\cal SP}H}=\Gamma_{\scriptscriptstyle{\cal SP}H}/
\bigl(\Gamma_h^{\rm SM}+\Gamma_{\scriptscriptstyle{\cal SP}H}\bigr)$\,
involving the rate $\Gamma_{\scriptscriptstyle{\cal SP}H}$ which is the sum of
$\Gamma(h\to\eta^{\scriptscriptstyle0}\eta^{\scriptscriptstyle0})$ or
$\Gamma(h\to\eta^{\scriptscriptstyle0}\nu N)$, depending on $m_0^{}$,
and~$\Gamma(h\to H\ell N)$.
The two numbers on each line in the ${\cal B}_{\scriptscriptstyle{\cal SP}H}$ column
correspond to the two numbers on the same line in the $\mu_2^{}$ column, where we have
included the possibility that $\mu_2^2$ can be negative~\cite{Deshpande:1977rw}.
Evidently, ${\cal B}_{\scriptscriptstyle{\cal SP}H}$ can be readily altered by
only varying~$\mu_2^{}$, with the other parameters fixed.

\begin{table}[t]
\caption{Sample values of mass parameters $m_{0,H}^{}$, $\mu_2^{}$, and $M_{1,2,3}$,
all in GeV, and Yukawa constants $Y_{1,2,3}$ satisfying the constraints discussed in
Section~\ref{constraints}.
The last three columns contain the resulting branching ratio
${\cal B}_{\scriptscriptstyle{\cal SP}H}$, in percent, of Higgs decay into final states containing
$\cal S, P$, or $H^\pm$ and ratio ${\cal R}_{\gamma{\cal V}^0}$ of
$\Gamma(h\to\gamma{\cal V}^{\scriptscriptstyle0})$ to its SM value for
\,${\cal V}^{\scriptscriptstyle0}=\gamma,Z$.\label{numbers}}
\small
\begin{tabular}{|cccccccccccc|} \hline
\,$m_0^{}$\, & \,$m_H^{}$\, & $\mu_2^{}$ & \,$M_1^{}$\, & \,$M_2^{}$\, & \,$M_3^{}$\, &
~ $Y_1$ \, & $Y_2$ & \, $Y_3$ ~ & ~~ ${\cal B}_{\scriptscriptstyle{\cal SP}H}\vphantom{\int_|^|}$ ~~
& ~ ~ ~ ${\cal R}_{\gamma\gamma}$ ~ ~ ~ & ~ ~ ~ ${\cal R}_{\gamma Z}$ ~ ~ ~ \\
\hline\hline
 50 &  70 &       46 (47)   &   9 &  13 &  63 & ~ 0.155 \, & \,0.372\, & \, 0.632 ~ & \,
 21 (14) \, &  0.89 (0.89) & 0.95 (0.95)\\
 60 &  80 &       54 (56)   &  10 &  15 &  71 & ~ 0.172 \, &   0.413   & \, 0.700 ~ &  26 (14) &
 0.91 (0.92) & 0.96 (0.97) \\
 70 &  70 &    $40i$ (25)   &  14 &  20 &  80 & ~ 0.155 \, &   0.378   & \, 0.677 ~ &  23 (11) &
 0.75 (0.83) & 0.88 (0.92) \\
 70 &  80 &      111 (99)   &  12 &  17 &  78 & ~ 0.180 \, &   0.437   & \, 0.730 ~ &  24 (12) &
 1.15 (1.09) & 1.06 (1.04) \\
 80 &  80 & \,159 $(88i)$\, &  15 &  22 &  82 & ~ 0.175 \, &   0.422   & \, 0.760 ~ &  22 (13) &
 1.53 (0.68) & 1.21 (0.86) \\
\, 120 \, & \, 70 \, & \,123 (111)\, & 20 & 29 & 85 & ~ 0.157 \, & 0.386 & \, 0.715 ~ & 20 (12) &
1.48 (1.34) & 1.20 (1.14) \\
\hline
\end{tabular}
\end{table}

The substantial values of ${\cal B}_{\scriptscriptstyle{\cal SP}H}$ in this table are made
possible by partial cancellations in $\Delta m_{ji}^2$ between the terms proportional to
$Y_{j,i}^4$ according to Eq.~(\ref{mass123}) and in the \,$\ell_j\to\ell_i\gamma$\, rates
between the different $Y_{\ell_i k}Y_{\ell_j k}$ terms with opposite signs, as Table~\ref{ym}
indicates, which allow $Y_{1,2,3}$ not to be too suppressed by the stringent constraints.
Some of such cancellations would not happen with solution sets of the type in Eq.~(\ref{332}).
We should mention, however, that the rather sizable $Y_{1,2,3}$ can result only
with some degree of fine-tuning, roughly at the per-mill level.
For such $Y_k$ and the masses in Table$\;$\ref{numbers}, we can reproduce the measured
$\Delta_{21,31}^2$ with \,$\lambda_5\sim2\times10^{-11}$.\,

To make comparison with LHC results, we find that the larger (unbracketed) numbers for
${\cal B}_{\scriptscriptstyle{\cal SP}H}$ in Table$\;$\ref{numbers} have begun to be probed
by bounds inferred from the latest Higgs measurements.
According to several analyses~\cite{h2inv}, the current data imply that the branching ratio
of nonstandard decays of the Higgs into invisible or undetected final-states can be as high as
22\% at 95\%\,CL if the Higgs production mechanism is~SM-like,
which is the situation in the scotogenic model.
These limits are not yet very strict and, as the bracketed
${\cal B}_{\scriptscriptstyle{\cal SP}H}$ numbers indicate, can be easily evaded by
changing~$\mu_2^{}$, which still has a~wide range of viability.
Therefore, the availability of other decay modes which may provide complementary
constraints would be highly desirable.

The impact of the new particles in the model can also be generated through loop
diagrams.\footnote{\baselineskip=11pt%
The new particles can contribute to $Z$-boson decays into neutrinos (charged leptons) via
one-loop diagrams with internal $\cal S$ or $\cal P$ $(H^\pm)$  and $N_k$.
At tree level, the $Z$ boson can also decay into three-body final states containing
them, similarly to the three-body Higgs decays above.
We have checked that these scotogenic effects on the $Z$ decay are not significant.}
Of great interest are their contributions to standard decay channels of the Higgs that are
already under investigation at the LHC.
Here we look at \,$h\to\gamma\gamma$\, and  \,$h\to\gamma Z$\, which arise in the SM mainly
from top-quark- and $W$-boson-loop diagrams and also receive one-loop contributions from~$H^\pm$.
These transitions are the same as those in the inert doublet model~\cite{Arhrib:2012ia,Cao:2007rm,idm}.
Based on general results in the literature~\cite{Djouadi:2005gi,Chen:2013vi},
the predicted rates of these modes are
\begin{eqnarray} \label{h2gg}
\Gamma(h\to\gamma\gamma) \,\,=\,\, \frac{\alpha^2 G_{\rm F}^{}\,m_h^3}{128\sqrt2\,\pi^3}
\Biggl|\frac{4}{3}\,A_{1/2}^{\gamma\gamma}\bigl(\kappa_t^{}\bigr)
+ A_1^{\gamma\gamma}\bigl(\kappa_W^{}\bigr)
+ \frac{m_H^2-\mu_2^2}{m_H^2}\, A_0^{\gamma\gamma}\bigl(\kappa_H^{}\bigr) \Biggr|^2 ~,
\end{eqnarray}
\begin{eqnarray} \label{h2gz}
\Gamma(h\to\gamma Z) &\,=&\,
\frac{\alpha G_{\rm F}^2\,m_W^2\bigl(m_h^2-m_Z^2\bigr)^3}{64\pi^4\,m_h^3} \Biggl|
\biggl(\frac{2}{c_{\rm w}^{}}-\frac{16s_{\rm w}^2}{3c_{\rm w}^{}}\biggr)
A_{1/2}^{\gamma Z}\bigl(\kappa_t^{},\lambda_t^{}\bigr)
+ c_{\rm w}^{}\,A_1^{\gamma Z}\bigl(\kappa_W^{},\lambda_W^{}\bigr)
\nonumber \\ && \hspace{23ex} - ~
\frac{\bigl(1-2s_{\rm w}^2\bigr)\bigl(m_H^2-\mu_2^2\bigr)}{c_{\rm w}^{}\,m_H^2}\,
A_0^{\gamma Z}\bigl(\kappa_H^{},\lambda_H^{}\bigr) \Biggr|^2 ,
\end{eqnarray}
where the expressions for the form factors $A_{1/2,1,0}^{\gamma\gamma,\gamma Z}$ are
available from~Ref.\,\cite{Chen:2013vi},
the $A_0^{\gamma\gamma,\gamma Z}$ terms originate
exclusively from the $H^\pm$ contributions, \,$\kappa_X^{}=4m_X^2/m_h^2$,\,
and~\,$\lambda_X^{}=4 m_X^2/m_Z^2$.\,
It is worth noting that in the \,$m_Z^{}=0$\, limit the amplitude for \,$h\to\gamma Z$\,
would reduce to that for \,$h\to\gamma\gamma$\, modulo the different $\gamma$ and $Z$
couplings to fermions, $W$ bosons, and~$H^\pm$.

In Table$\;$\ref{numbers} we have also listed the resulting numbers for the ratio
\begin{eqnarray}
{\cal R}_{\gamma{\cal V}^{\scriptscriptstyle0}}^{} \,\,=\,\,
\frac{\Gamma(h\to\gamma{\cal V}^{\scriptscriptstyle0})}
{\Gamma(h\to\gamma{\cal V}^{\scriptscriptstyle0})_{\rm SM}^{}} ~, \hspace{5ex}
{\cal V}^{\scriptscriptstyle0} \,\,=\,\, \gamma, Z ~,
\end{eqnarray}
where $\Gamma(h\to\gamma{\cal V}^{\scriptscriptstyle0})_{\rm SM}^{}$ is the SM rate,
without the $A_0^{\gamma{\cal V}^0}$ part.
The examples in this table demonstrate that the scotogenic effects on
$\Gamma(h\to\gamma\gamma)$ and $\Gamma(h\to\gamma Z)$ have a positive
correlation.
Comparing the ${\cal B}_{\scriptscriptstyle{\cal SP}H}$ and
${\cal R}_{\gamma\gamma,\gamma Z}$ numbers, we see that the latter have the milder
dependence on~$\mu_2^{}$, unless $\mu_2^2$ changes signs.
Moreover, there does not appear to be a clear correlation between
${\cal B}_{\scriptscriptstyle{\cal SP}H}$ and the impact of $H^\pm$
on~$\Gamma(h\to\gamma{\cal V}^{\scriptscriptstyle0})$, which is partly due to the fact that
the Yukawa parameters $Y_k$ which are present in $\Gamma_{\scriptscriptstyle{\cal SP}H}$ do not
contribute to $\Gamma(h\to\gamma{\cal V}^{\scriptscriptstyle0})$.
Since \,$h\to\gamma\gamma$\, has been detected, unlike the $\gamma Z$ mode~\cite{h2gz},
we can already compare our examples with the data.
The latest measurement of the signal strength for \,$h\to\gamma\gamma$\, performed by
the ATLAS Collaboration is~\,$\sigma/\sigma_{\rm sm}^{}=1.6\pm0.3$\,~\cite{atlas}.
On the other hand, for the same mode the CMS Collaboration has found
\,$\sigma/\sigma_{\rm SM}^{}=0.78\pm0.27$\, and \,$1.11\pm0.31$\,
using two different methods~\cite{cms:h2gg}.
While awaiting an experimental consensus on this decay channel, we can say that all of
the ${\cal R}_{\gamma\gamma}$ numbers in Table$\;$\ref{numbers} are still compatible with
one or more of these LHC results, but the situation will likely change when
more data become available.
We may also expect that supplementary information will be supplied by future observations
of~\,$h\to\gamma Z$.\,

We have seen that there are some decay modes of the Higgs boson that can be employed
to test the scotogenic model in complementary ways.
At this point the restrictions inferred from the existing data on the Higgs decays are
not yet strong, but already start to probe the parameter space allowed by other data.
In addition, with the fairly large Yukawa couplings which we have obtained and the relatively
light charged scalars, direct searches at the LHC may offer extra
tests~\cite{Bhattacharya:2013mpa} on the scenario treated here.

\section{Conclusions\label{summary}}

We have explored for the scotogenic model of radiative neutrino mass
some implications of the recent discovery of a Higgs boson at the LHC
and experimental determination of $\sin\theta_{13}$ that is not very small.
Employing a two-parameter neutrino-mixing matrix which is consistent with the latest
neutrino-oscillation data within one sigma, we derive solutions for the Yukawa couplings
of the nonstandard particles in the model, which consist of scalars and fermions.
Such solutions are also applicable to some other models of radiative neutrino mass.
We select one of the new fermions to be the lightest of the nonstandard particles
which plays the role of cold~DM.
Subsequently, taking into account various constraints, including those from low-energy
measurements and the observed relic density, we use the solutions for the Yukawa couplings
to consider Higgs decays into final states containing the new particles.
We find that within the allowed parameter regions the rates of such decays can be
significant, which can already be probed with the latest Higgs measurements.
We also examine how these exotic decay channels may correlate with the scotogenic effects on
the Higgs decays into $\gamma\gamma$ and $\gamma Z$, which are under intensive study at the LHC.
Consequently, upcoming Higgs data with improved precision from the LHC, or a future Higgs
factory, can be expected to reveal hints of the new particles or impose further restrictions
on the model.

\acknowledgments

We would like to thank C.Q. Geng, X.G. He, and M. Kohda for helpful discussions.
This work was supported in part by NCTS.

\appendix

\section{Solutions for Yukawa couplings $\bm{{\cal Y}_{jk}}$\label{yjk}}

One can solve the diagonalization conditions in Eq.\,(\ref{diagcond}) exactly for
the three pairs of Yukawa couplings $(Y_{ek},Y_{\mu k})$, \,$k=1,2,3$,\,
in terms of \,$Y_k=Y_{\tau k}$.\,
There is more than one set of the solutions.
In each set, we can express the pairs as
\,$(Y_{ek},Y_{\mu k})=\bigl(\bar e_z^{},\bar\mu_z^{}\bigr)Y_k^{}$,\, where \,$z=a,b$, or $c$\, and
\begin{eqnarray} \label{em} & \displaystyle
\bar e_a^{} \,\,=\,\, \frac{\sqrt2\;c_\theta^{}\,c_\varsigma^{}}
{s_\theta^{}\,c_\varsigma^{}-s_\varsigma^{}} ~, \hspace{5ex}
\bar e_b^{} \,\,=\,\, \frac{-\sqrt2\;s_\theta^{}}{c_\theta^{}} ~, \hspace{5ex}
\bar e_c^{} \,\,=\,\,
\frac{\sqrt2\;c_\theta^{}\,s_\varsigma^{}}{s_\theta^{}\,s_\varsigma^{}+c_\varsigma^{}} ~, &
\nonumber \\ & \displaystyle
\bar\mu_a^{} \,\,=\,\, \frac{s_\varsigma^{}+s_\theta^{}\,c_\varsigma^{}}
{s_\varsigma^{}-s_\theta^{}\,c_\varsigma^{}} ~, \hspace{7ex}
\bar\mu_b^{} \,\,=\,\, -1 ~, \hspace{8ex}
\bar\mu_c^{} \,\,=\,\, \frac{c_\varsigma^{}-s_\theta^{}\,s_\varsigma^{}}
{s_\theta^{}\,s_\varsigma^{}+c_\varsigma^{}} ~. &
\end{eqnarray}
Since two or all three of the pairs may share the same~$z$, such as in Eq.\,(\ref{332}),
there are altogether 27 sets of the solutions to~Eq.\,(\ref{diagcond}).
Not all of them can lead to at least two nonzero masses  among the eigenvalues
in~Eq.\,(\ref{numass}).
Three of the sets can each only yield one nonzero mass, whereas 18 (six) of the others can
lead to two (three) nonzero masses.

\section{Dark matter annihilation amplitudes\label{Mann}}

The Majorana nature of $N_j$ implies that the process \,$N_k\bar N_l\to\ell_i^-\ell_j^+$\,
arises from $H$-mediated $t$- and $u$-channel diagrams.
We find the absolute square of its amplitude, averaged (summed) over initial (final) spins,
to be
\begin{eqnarray} \label{NN2ll}
\overline{\bigl|{\cal M}_{N_k\bar N_l\to\ell_i^-\ell_j^+}\bigr|^2} &\,=&\,
|{\cal Y}_{ik\,}{\cal Y}_{jl}|^2\,
\frac{\bigl(M_k^2+m_{\ell_i}^2-t\bigr)\bigl(M_l^2+m_{\ell_j}^2-t\bigr)}{4\bigl(m_H^2-t\bigr)^2}
\nonumber \\ && + ~
|{\cal Y}_{il\,}{\cal Y}_{jk}|^2\,
\frac{\bigl(M_k^2+m_{\ell_j}^2-u\bigr)\bigl(M_l^2+m_{\ell_i}^2-u\bigr)}{4\bigl(m_H^2-u\bigr)^2}
\nonumber \\ && + ~
{\rm Re}\bigl({\cal Y}_{ik\,}^*{\cal Y}_{jl\,}^{}{\cal Y}_{il\,}^{}{\cal Y}_{jk}^*\bigr)\,
\frac{M_k^{}M_l^{}\, \bigl(m_{\ell_i}^2+m_{\ell_j}^2-s\bigr)}
{2\bigl(m_H^2-t\bigr)\bigl(m_H^2-u\bigr)} ~,
\end{eqnarray}
where
\begin{eqnarray}
s \,\,=\,\, \bigl(p_{N_k}+p_{N_l}\bigr)^2 ~, \hspace{5ex}
t \,\,=\,\, \bigl(p_{N_k}-p_{\ell_i}\bigr)^2 ~, \hspace{5ex}
u \,\,=\,\, \bigl(p_{N_k}-p_{\ell_j}\bigr)^2 ~.
\end{eqnarray}
In the case of \,$N_k\bar N_l\to\nu_i^{}\bar\nu_j^{}$,\, proceeding from
$({\cal P,S})$-mediated $t$- and $u$-channel diagrams, we need to take into account
the Majorana nature of the final neutrinos as well.
It follows that for \,$m_0^{}\simeq m_{\cal S}^{}\simeq m_{\cal P}^{}$\,
and negligible $\nu$ masses
\begin{eqnarray} \label{NN2nn}
\overline{\bigl|{\cal M}_{N_k\bar N_l\to\nu_i\bar\nu_j}\bigr|^2} &\,=&\,
|{\cal Y}_{ik\,}{\cal Y}_{jl}|^2\,\frac{\bigl(M_k^2-t\bigr)\bigl(M_l^2-t\bigr)}
{4\bigl(m_0^2-t\bigr)^2}
\,+\,
|{\cal Y}_{il\,}{\cal Y}_{jk}|^2\,\frac{\bigl(M_k^2-u\bigr)\bigl(M_l^2-u\bigr)}
{4\bigl(m_0^2-u\bigr)^2}
\nonumber \\ && - ~
{\rm Re}\bigl({\cal Y}_{ik\,}^*{\cal Y}_{jl\,}^{}{\cal Y}_{il\,}^{}{\cal Y}_{jk}^*\bigr)\,
\frac{M_k^{}M_l^{}\,s}{2\bigl(m_0^2-t\bigr)\bigl(m_0^2-u\bigr)} ~,
\end{eqnarray}
where now \,$t=\bigl(p_{N_k}-p_{\nu_i}\bigr)^2$\, and \,$u=\bigl(p_{N_k}-p_{\nu_j}\bigr)^2$.\,

\end{document}